\documentclass[%
 reprint,
superscriptaddress,
 amsmath,amssymb,
 aps,
]{revtex4-1}

\usepackage{color}

\usepackage{graphicx}
\usepackage{dcolumn}
\usepackage{bm}

\begin{document}


\title{Motif Dynamics in Signed Directional Complex Networks}

\author{Youngjai Park}
 \affiliation{%
 Department of Applied Physics, Hanyang University, Ansan 15588, Korea
}%
\author{Mi Jin Lee}
\email{mijinlee@hanyang.ac.kr}
 \affiliation{%
 Department of Applied Physics, Hanyang University, Ansan 15588, Korea
}%
\author{Seung-Woo Son}%
 \email{sonswoo@hanyang.ac.kr}
 \affiliation{%
 Department of Applied Physics, Hanyang University, Ansan 15588, Korea
}%
\affiliation{%
 Asia Pacific Center for Theoretical Physics (APCTP), Pohang 37673, Korea
}%

\date{\today}

\begin{abstract}

Complex networks evolve and vary their structure as time goes by. In particular, the links in those networks have both a sign and a directionality. To understand their structural principles, we measure the network motifs, which are patterns that appear much more than one would expect in randomized networks, considering both link properties. We propose motif dynamics, which is a study to investigate the change in the number of motifs, and applied the motif dynamics to an open evolving network model and empirical data. We confirm that a non-cyclic motif has a greater correlation with the system size than a cyclic structural motif. Furthermore, the motif dynamics can give us insight into the friendship between freshmen in a university.

\begin{description}
\item[Keyworks]
Network motifs, Signed directional complex networks, Temporal networks
\end{description}
\end{abstract}

\maketitle


\section{Introduction}
\label{sec:introduction}

Complex systems such as brain, ecosystem, financial market, and our society continue to undergo change in time with interacting with their surrounded environment~\cite{paranjape2017motifs,li2017fundamental,cencetti2020temporal}. It can be considered as an open evolving complex network, where a node is each neuron, species, or individual, and a connecting link indicates a signal path, a biomass flow, or an interaction between paired nodes. Depending on the types of interactions, one can assign the link additional attributes such as direction (incoming/outgoing), weight (intensity of interaction), and sometimes with its $+/-$ sign~\cite{benson2016higher,peel2018multiscale,kirkley2019balance}. The complicated interactions described with the various types of links give rise us to the intriguing emergent phenomena such as a formation of the giant component, which corresponds to the outbreaks in epidemics and neuronal avalanches in our brain.

Starting from investigation of basic properties such as the (in-/out-)degree distribution, weight distribution, and averaged quantities of interest, many researchers have tried to understand the emergent patterns. Usually, the investigations have been undertaken at a macroscopic level. Recently, the understanding at a microscopic scale also have been emphasized as well, because eventually the accumulated effect of the local structure (small interaction pattern) brings the macroscopic observation. As one of the ways to figure out the local structure, the network motif analysis is suggested~\cite{itzkovitz2003subgraphs,prill2005dynamic,grilli2017higher}. The motifs in network science are statistically significant subnetworks, which contain specific interaction patterns among the small number of nodes. For instance, let us consider the triplet nodes in a network with undirectional (symmetric) links. Then, the possible motifs in the triplet are an open and a closed triangle with two and three links, respectively. The frequency of the motif of a triplet is often estimated by the clustering coefficient, the well-known global measurement of networks~\cite{luce1949method,watts1998collective,onnela2005intensity}. The motif is naturally extended to the directional (asymmetric) and signed link even though the handling of complexity incredibly increases.

In this paper, we take account into a link with a direction (asymmetric relation) as well as a $+/-$ sign (positive or negative influence) in an open evolving system, motivated by the ecological system~\cite{shimada2014universal}. In this system, the meaning of a link completely varies with the sign. In the past decade, a number of studies have considered only the direction or only the sign with links in various fields for motif analysis~\cite{milo2002network,sporns2004motifs,alon2007network,facchetti2011computing}. Since there exists the difference of roles between a positive and negative link, we count in both the direction and sign of links. By inspecting the frequency or abundance of each motif in a network, we are aiming at understanding what interaction patterns mainly contribute to the emergent behaviors. Furthermore, the motif abundance is also traced as time goes by, since we consider the open evolving system where the interaction patterns change in time. We call the temporal change a \textit{motif dynamics}. That is, we try to figure out the emergent pattern of open evolving networks in terms of the motif dynamics~\cite{kim2014dynamic,kim2014network}.

To analyze the motif dynamics, we first reduce all the possible motifs to 22 types consisting of three nodes with only uni-directional (asymmetric) links. To evaluate the significance of the motif abundance, we compute the $Z$-score, and then the motif dynamics is characterized by the temporal behavior of $Z$-score. We apply this setup to both a theoretical model network and empirical data. The model network is of mimicking the ecological system~\cite{murase2010simple,shimada2014universal,murase2015universal,ogushi2017enhanced,murase2018conservation,ogushi2019temporal}, and the empirical data are collected from KONECT~\cite{konect}, which contains many kinds of network data sets in a variety of domains. We find that particular motifs mainly devote to the network growth, and that the motif dynamics can be interpreted in various contexts depending on networks. We expect that analysis with the motif dynamics helps us to understand better our surrounding complex systems. Furthermore, this approach provides a new insight into complex systems management by considering signs of interactions.

This paper is organized as follows: The explanations of the motif and $Z$-score are portrayed in Sec.~\ref{sec:methods}. In Sec.~\ref{subsec:modelnetwork}, motif dynamics on an open evolving model are described. $Z$-scores of non-cyclic and cyclic motifs are compared to test the non-cyclic dominance in empirical data in Sec.~\ref{subsec:noncyclic}. Section~\ref{subsec:friendship} is focused on a friendship network data among university freshmen and its motif dynamics. Further discussion is in Sec.~\ref{sec:conclusions}.

\section{Methods}
\label{sec:methods}

A network motif is a frequently observed subnetwork pattern in a given signed directional network $G(N, L)$ with $N$ nodes and $L$ links. The motif set $\mathbb{M}(n, l)~\subseteq~G(N, L)$ is a set of motifs having $n < N$ nodes and $l < L$ links. We are not concerned with all the motifs, but specific and contributive ones to the system. To evaluate it, we compute the expected appearance of a motif $m \in \mathbb{M}(n, l)$ compared with those in the randomized networks, namely, the $Z$-score of the frequency or abundance $Z(m)$ defined as
\begin{equation}
    Z(m)=\frac{O(m)-\bar{O}_{\rm {null}}(m)}{\sigma_{\rm {null}}(m)},
\label{eq:zscore}
\end{equation}
where $O(m)$ is the number of the observation/appearance (abundance) of a specific motif $m$ in a given original network, and $\bar{O}_{\rm {null}}(m)$ and $\sigma_{\rm {null}}(m)$ are the average abundance and its standard deviation in the randomized networks, respectively. Thus, Eq.~(\ref{eq:zscore}) means how significantly the motif $m$ is observed comparing to other networks which have the same number of nodes and links. The utilization of the $Z$-score also helps us to compare the tendency of the motif abundance across different networks with different sizes at a different time.

\begin{figure}
\centering
\includegraphics[width=0.85\linewidth]{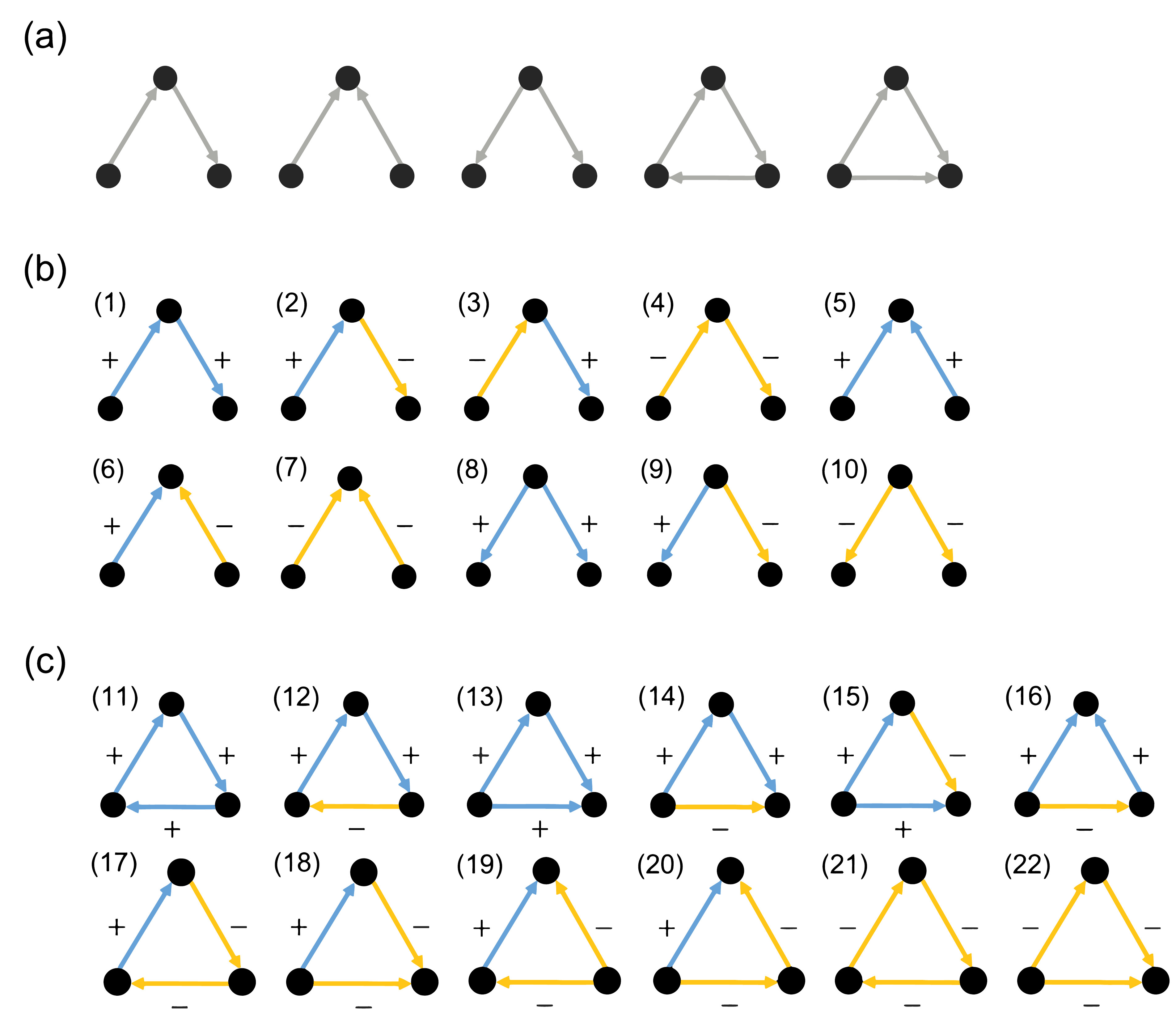}
\caption[All the possible motifs with three nodes]{All the possible motifs with three nodes. (a) There exist 5 unsigned directional motifs, and 22 signed directional motifs composed of 10 motifs and 12 motifs belonging to $\mathbb{M}(3, 2)$ and $\mathbb{M}(3, 3)$. The blue line represents a positive ($+$) influence on a target, and the orange one indicates a negative ($-$) influence.
}
\label{fig:motifs}
\end{figure}

Motifs provide insight of important functional structure in a system, but detecting the important structure involves the computationally challenging problem. Moreover, we consider the direction and sign of the link, so the cardinality of $\mathbb{M}(n, l)$ becomes even larger than in a network with either signed or directional links. For simplicity, we pay attention to open or closed triplets, i.e., $\mathbb{M}(3, 2)$ and $\mathbb{M}(3, 3)$. The triplets play roles as fundamental building blocks in constituting the system. While constructing the sets, the bidirectional links are excluded because there are 132 types of motifs if we consider the bidirectional links. It is unlikely to understand all the meaning of 132 motifs. Following this criteria, we collect 22 signed directional motifs as shown in Figs.~\ref{fig:motifs}(b) and~\ref{fig:motifs}(c). If one takes only the direction apart from the sign into account, there are only five motifs considering the mirror and rotational symmetry [Fig.~\ref{fig:motifs}(a)]. The expanded number of the motifs means that the effective role of a link varies depending on its sign even on the same connectivity structure as in an unsigned case. For example, the feedback loop of positive links with three nodes makes the structure robust. However, a negative feedback loop with three nodes can cause the collapse of the system, despite the same structure if we only consider the direction~\cite{alon2007network}.

To obtain the statistical significance of the motif abundance, we need a null model to compare in Eq.~(\ref{eq:zscore}) constructing randomized networks by rewiring links. The rewiring process in the signed and directional network is based on the proposed method to avoid self-/multi-links~\cite{milo2002network,iorio2016efficient}. One can perform the rewiring process considering four conservation conditions: i) nothing to keep, ii) degree as in undirectional networks, iii) in-/out-degree, or iv) in-/out-degree with sign. In this study, to inspect the signed directional motifs, we implement the rewiring with keeping the sign and directionality sequence for every node in order to compare the original network with randomized networks under the same macroscopic conditions.

\begin{figure*}
\centering
\includegraphics[width=0.6\linewidth]{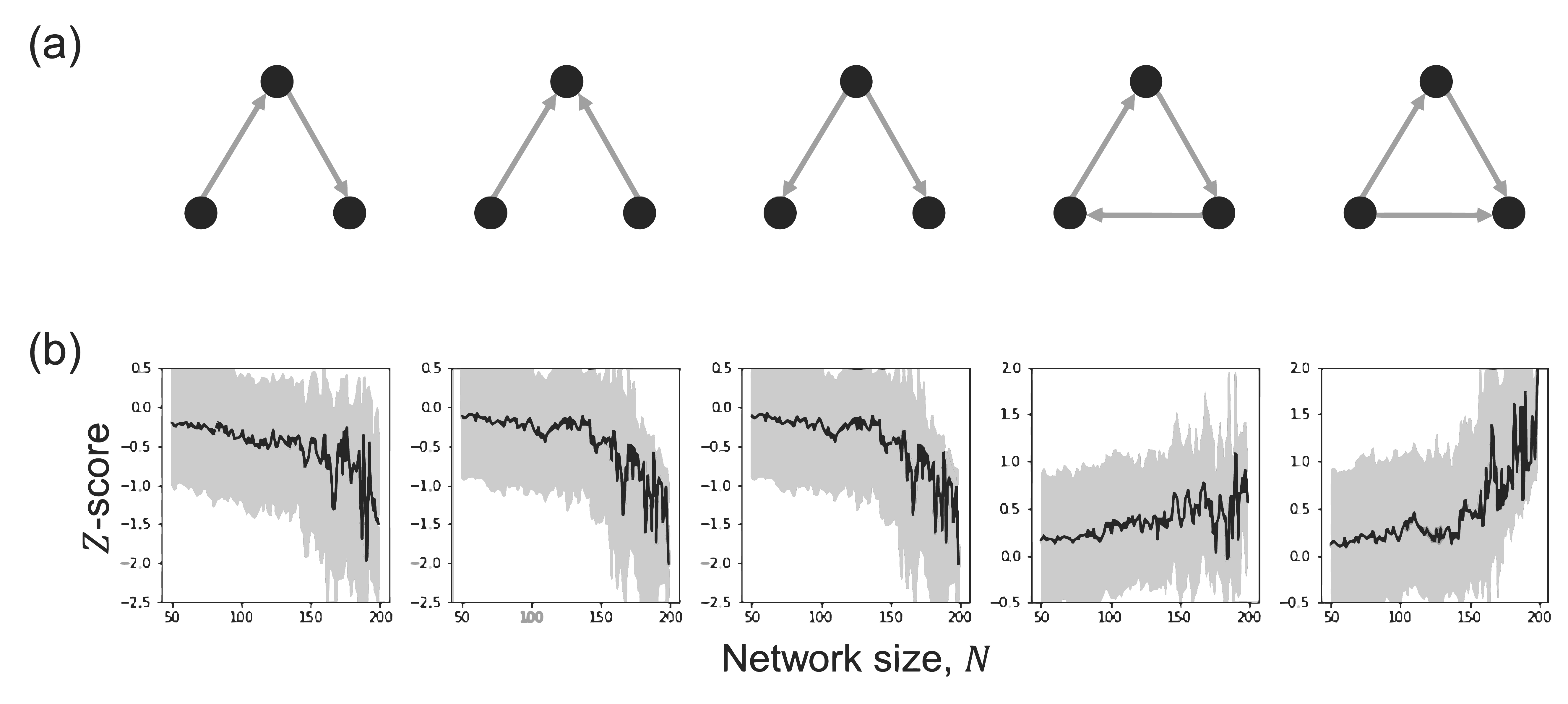}
\caption[Unsigned motif dynamics in the model]{Motif dynamics over the network size. (a) The five types of unsigned directional motifs consist of three open triangle types and two closed triangle types when the model grows up at a diverging phase. (b) Each panel displays the motif dynamics. The solid line represents the ensemble average and the shading area indicates its standard deviation. In the open evolving model, the $Z$-scores of closed triangles increases as the network grows, but those of open triangles decrease.}
\label{fig:OEN_unsigned}
\end{figure*}

\section{Results}
\label{sec:results}
\subsection{Open evolving network model}
\label{subsec:modelnetwork}

Multiple species in an ecosystem have influence on or get influenced by each other in positive, negative, or neutral ways, that leads to which species survive in the system. According to a relationship between two different species based on the influence, several typical relationships are observed in the ecosystem such as competition, mutualism, and neutralism~\cite{lidicker1979clarification,sahney2010links}. 
This influenceable structure across species is represented as a complex network. Every single species corresponds to a node, and the interaction between them is described as a link. In particular, the signed directional link is adopted to describe the influence from/on species around a node of interest in a positive/negative/neutral way. $w_{ij}>0$ ($<0$) denotes the positive (negative) influence of a node $i$ to a node $j$, and $w_{ij}\neq w_{ji}$ in general.

The total influence exerted on a species (node) $i$ is relevant to its fitness and survivability. It is constantly changing in time as the system experiences the extinction of existing species and introduction of invasive ones. To portray the situation, an open evolving network model has been proposed~\cite{shimada2014universal}. The total influence is defined as a fitness of node $i$ given by $f_i=\sum_{j} w_{ji}$, where $w_{ji}$ is an weighted incoming edges in the viewpoint of node $i$. In this model, the weight $w_{ji}$ is randomly drawn from the normal distribution, and the direction is also assigned at random. A new invasive species with $k$ links is introduced into the system every time step and interacts with the $k$ randomly selected resident species. A species with $f_i \leq 0$ goes extinct, and the fitness of species connected with the extinct one is recalculated. The extinction and recalculation iterate until no more extinction occurs. We adopt the aforementioned model and generate synthetic ecosystems. One of the important results in this model is an emergence of the diverging phase in terms of system size, in $5\lesssim k \lesssim 18$. Since large networks are convenient to analyze statistically, $k=10$ models are investigated.

\begin{figure*}
\centering
\includegraphics[width=0.6\linewidth]{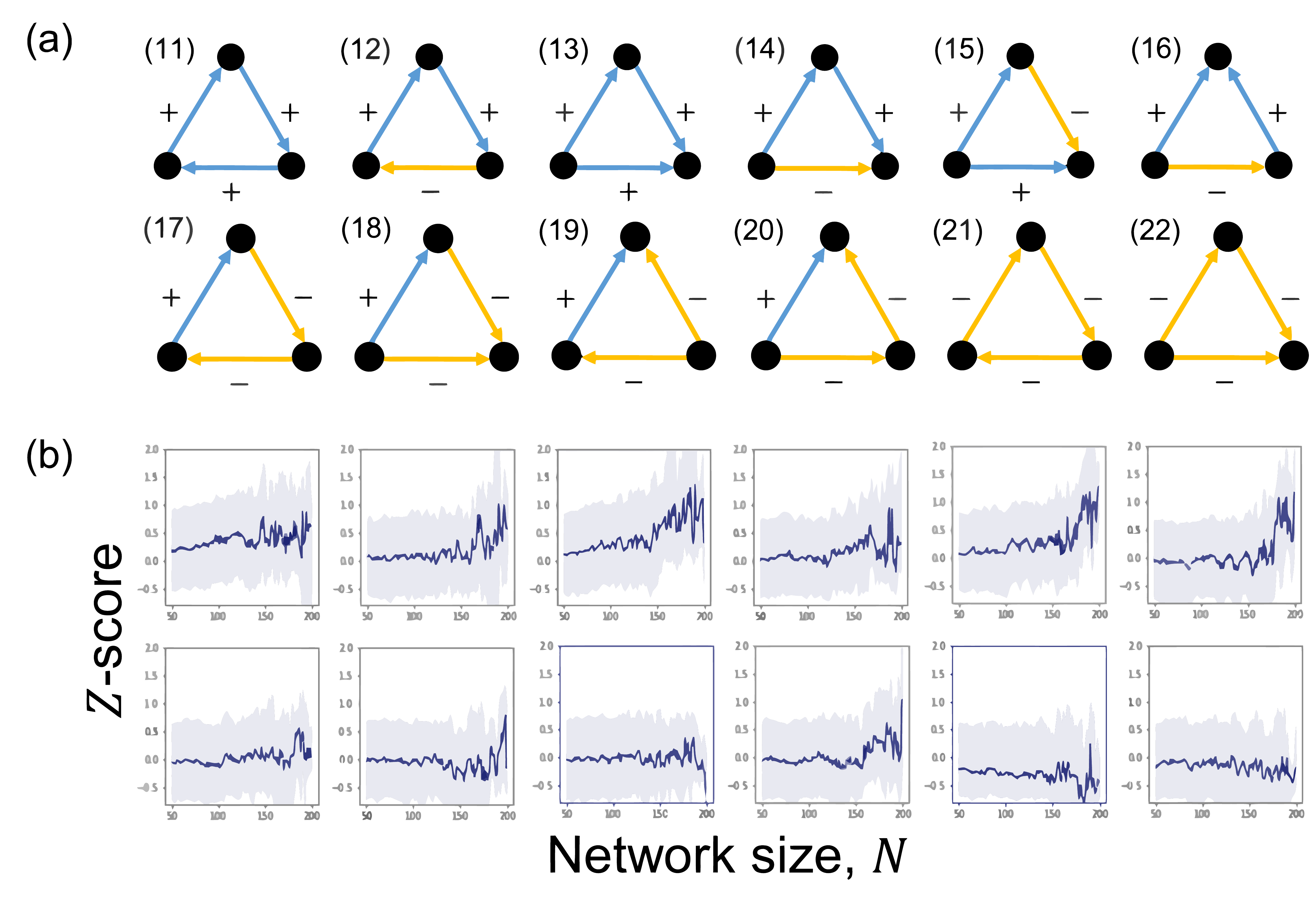}
\caption[Signed motif dynamics in the model]{Singed motif dynamics in the model. (a) 12 closed triangle motifs with sign, $\mathbb{M}(3,3)$. (b) Each panel displays the $Z$-score change as the network size increases. The solid line represents the ensemble average and the shading area indicates its standard deviation. The motifs in the upper row of (b), which contain more positive links, become abundant as network grows.}
\label{fig:OEN_signed}
\end{figure*}

We collect the motifs belonging to $\mathbb{M}(3, 2)$ and $\mathbb{M}(3, 3)$ in Figs.~\ref{fig:motifs}(b) and~\ref{fig:motifs}(c) of the synthetic ecosystems. We analyze the motif dynamics by means of $Z$-score in Eq.~(\ref{eq:zscore}) and which types of interaction structure mainly contribute to the diverging phase. During the simulation, the network size $N(t)$ linearly increases as time $t$, so $N(t)$ is nearly equivalent to $t$ in Figs.~\ref{fig:OEN_unsigned} and~\ref{fig:OEN_signed}. Figure~\ref{fig:OEN_unsigned} shows the results of unsigned directional motifs and signed directional motifs are displayed in Fig.~\ref{fig:OEN_signed}.

As a first step to understand the motif dynamics, we inspect the simple unsigned directional motifs in Fig.~\ref{fig:OEN_unsigned}. The unsigned directional motifs are combinations of the signed directional ones. For example, the motifs 1, 2, 3, and 4 labeled in Fig.~\ref{fig:motifs}(b) degenerate into the first leftmost one in Fig.~\ref{fig:motifs}(a), and then $Z$-score in Fig.~\ref{fig:OEN_unsigned}(b) is calculated from the randomization conserving only in-/out-degree. The abundance of a motif changes in a network size and link density. Interestingly, distinct behaviors of open and closed triplets are observed in our simulation. As the network grows, the closed ones are discovered more and more, contrasted with the remarkably decreasing tendency of the open triplets as shown in Fig.~\ref{fig:OEN_unsigned}(b). One may think that the large $k$ makes the dense connectivity in the network. In this model, however, the large $k$ does not guarantee the creation of the closed triplet because the node deletion (extinction of species) frequently occurs at $k \gtrsim 18$, which leads to a bunch of the removal of the associated links. Hence, the increasing tendency of closed triplets is a nontrivial finding.

From the fact that the unsigned directional closed-triplet is correlated with the growth of the giant component, the closed triplet is worth to be investigated in detail. The two motifs of the closed one in Fig.~\ref{fig:OEN_unsigned}(a) are subdivided into twelve motifs $m\in \mathbb{M}(3, 3)$ when the sign is considered as in Fig.~\ref{fig:OEN_signed}(a). Interestingly, the increasing pattern observed in Fig.~\ref{fig:OEN_unsigned}(b) is not always observed: some motifs appear significantly, and others do not. Usually, the motifs with more than two positive-signed links (i.e., the majority) tend to show the notable increasing pattern and large value of $Z$-score in large $N$ [see the panels in the respective upper line in Figs.~\ref{fig:OEN_signed}(a) and~\ref{fig:OEN_signed}(b)]. The motif 13 and 22 have the same structural feature except for a sign imposed on links, but their motif dynamics are remarkably different. The result of the different tendencies by the fraction of positive links seems quite reasonable. Owing to the model characteristic that a species having $f_i >0$ survives, the negative link seldom exists. In order to make networks grow without shrinkage, the extinction (node deletion) has to rarely happen, and then the least negative links are beneficial to system. Unlike the case of $\mathbb{M}(3, 3)$, we confirm that all the motifs $m \in \mathbb{M}(3, 2)$ maintain the decreasing tendency (not shown in this paper).

\subsection{Non-cyclic motif in the networks}
\label{subsec:noncyclic}

In Sec.~\ref{subsec:modelnetwork}, there is one more interesting finding that the feedforward motif (motif 13) has a greater correlation with $N$ than the feedback structure (motif 11) [Fig.~\ref{fig:OEN_signed}]. The two motifs have the positive links only in common but the different structure in terms of directionality as cyclic and non-cyclic structures for the motif 11 and 13, respectively. It is probably caused from the vulnerability against node deletion which is inherent in the feedback structure like the motif 11. When one node disappears, the fitness of the other nodes decreases in the cyclic structure, leading to a cascading failure. In contrast, it is well known that the non-cyclic motif is stable against external fluctuations~\cite{park2020cyclic}.

\begin{figure}[b]
\centering
\includegraphics[width=0.85\linewidth]{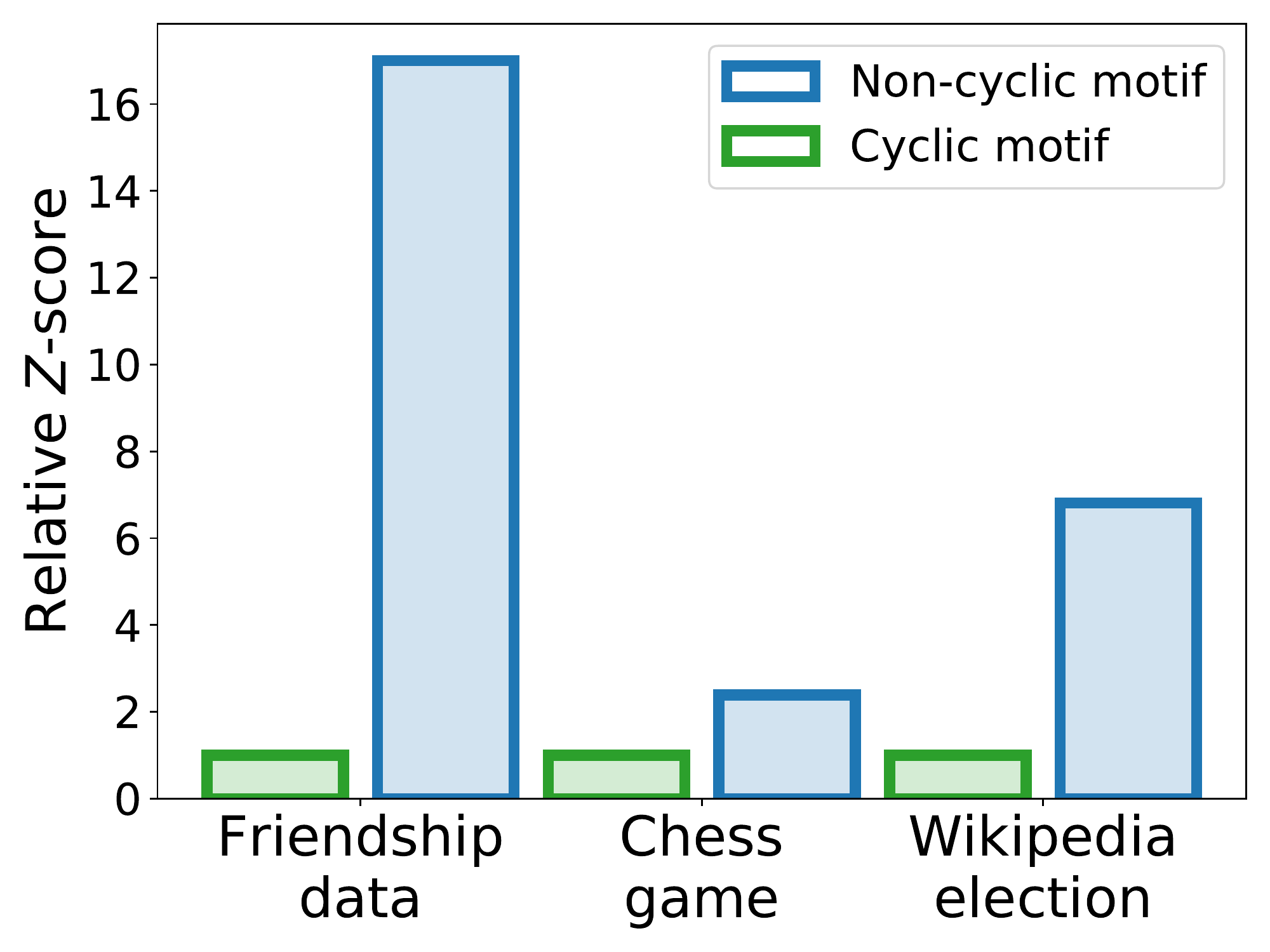}
\caption[Non-cyclic and cyclic motif in networks]{The relative $Z$-scores of cyclic (green) and non-cyclic (blue) motifs in the three empirical network data. The non-cyclic motif appears more frequently in all the three as well as a model network. In order to represent the relative proportions, we examine the relative $Z$-score normalizing by the $Z$-score of motif 13.}
\label{fig:non_cyclic}
\end{figure}

\begin{figure*}
\centering
\includegraphics[width=0.6\linewidth]{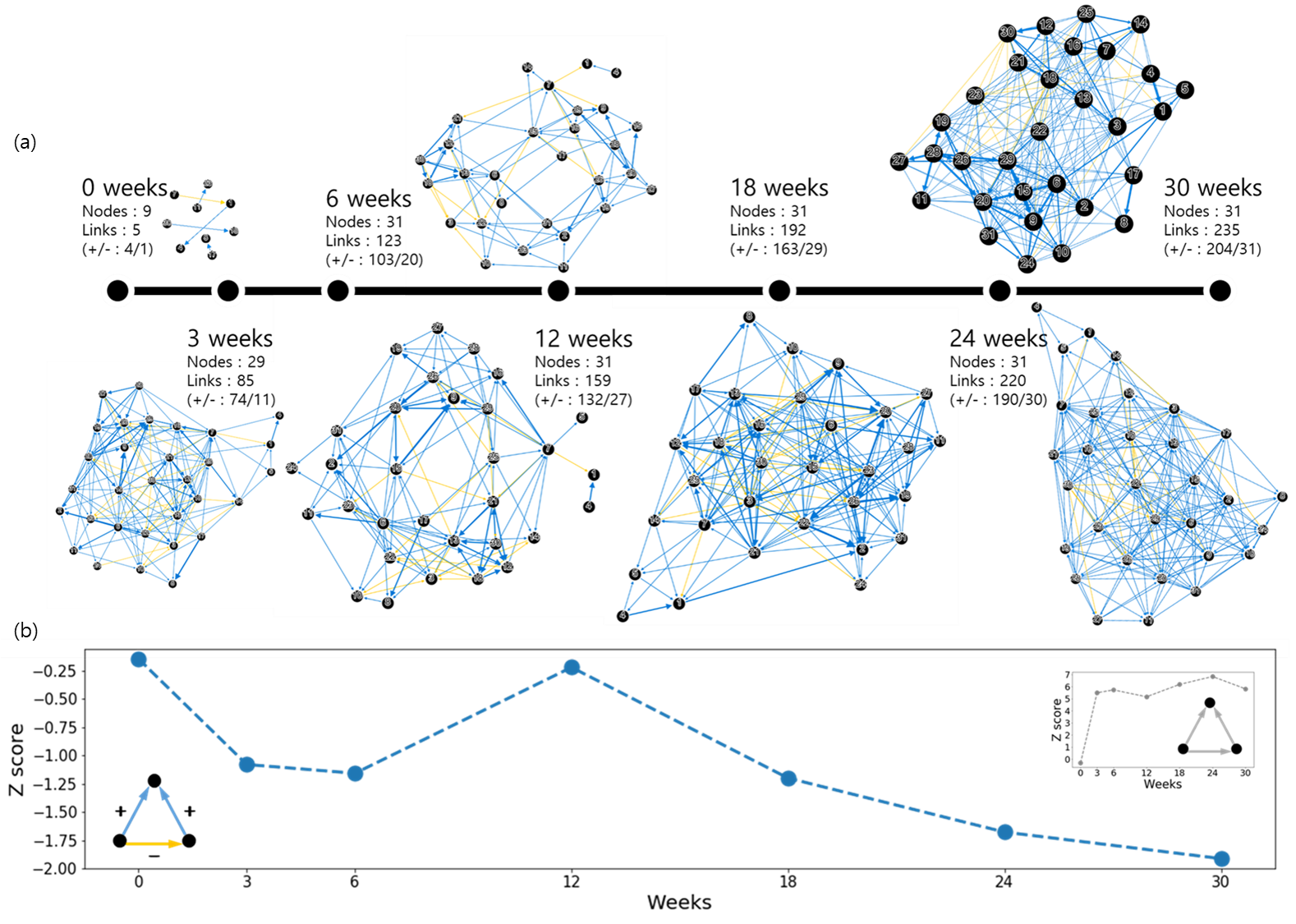}
\caption[Friendship network on a university data]{The $Z$-score change of a motif 16 in a friendship network among university freshmen in the Netherlands as time goes by. (a) The visualization of the network data in a timeline: nodes mean the university freshmen and links represent the relationship between two students depending on their rating score (positive: blue, negative: orange). (b) $Z$-score change of the motif 16 as time goes by: there is a rise and fall like the dramatic structure (exposition, inciting incident, rising action, climax, falling action, resolution, and denouement). At the twelve week, the conflict structure (motif 16) rise high and the $Z$-score goes down around the end of the semester. The inset is about the motif if we do not consider the sign. It is insufficient to explain the conflict among the freshmen.}
\label{fig:dutch_univ}
\end{figure*}

We confirm that the motif significance of the non-cyclic structure (motif 13) is higher than that of the cyclic structure (motif 11) by examining and comparing the $Z$-score even in empirical data. We utilze data provided from the KONECT dataset~\cite{konect}: ``Friendship in a university'' (we call it ``Friendship data'')~\cite{van1999friendship}, ``Chess game''~\cite{konect:2016:chess}, and ``Wikipedia election''~\cite{konect:2016:elec,konect:leskovec207}. In the ``Friendship data'', nodes are freshmen in a university in the Netherlands, and links are rating scores to their classmates. The rating score ranges from -1 to 3 (see in Sec.~\ref{subsec:friendship} in more detail). In the ``Chess game'', a node represents a player, and the result of a match between two players is a link. The direction of the link is defined as white-to-black since white moves first, and the weight is expressed as 1 if white wins, 0 if draw, and -1 if white loses. The ``Wikipedia election’’ contains the voting records of an administrator's election in Wikipedia. A node represents a voter, and a link is marked as 1 (-1) with a vote for (against) the candidate.

For comparison, we plot the relative $Z$-score in Fig.~\ref{fig:non_cyclic}. One can see that the non-cyclic motif has the higher significance than that of the cyclic motif as expected. The significant observation of the non-cyclic motif may be one of the indices of the robustness of the system.

\subsection{Friendship network among university freshmen}
\label{subsec:friendship}

We additionally analyze the meaning of the motif dynamics in the context of the empirical data. To do so, we use the ``Friendship data'' introduced in Sec.~\ref{subsec:noncyclic}. A university in the Netherlands conducted a survey of freshmen during 1994 and 1995~\cite{van1999friendship}. In the survey, freshmen were asked to rate their friendship except for themselves with the integer-valued score ranging from -1 to 3. From the response to the survey, the friendship network is constructed with a node and a link being a freshman and a rating score to a friend, respectively. The network satisfies both the interaction type (social interaction) and the form (the temporal network with the signed weight) similarly to those of the ecological model network in Sec.~\ref{subsec:modelnetwork}.

The survey conducted 7 times starting from the beginning of a semester during the given period with the different intervals. As time goes, the giant component is formed with a larger size because more students participate in the survey as seen in Fig.~\ref{fig:dutch_univ}(a). A meaning of a motif is interpreted in various ways depending on a context of a network. We put focus on the motif 16 in Fig.~\ref{fig:motifs}(c), because the structure implies a conflict relationship among three students: two students like the other student simultaneously, but one student of the two dislikes the other of the two. We try to understand the meaning of the motif dynamics, so only concentrate on analyzing the dynamics of the motif 16 of interest rather than all the motifs in $\mathbb{M}(3, 2)$ and $\mathbb{M}(3, 3)$.

The dynamics of the motif 16 shows a quite interesting pattern as seen in Fig.~\ref{fig:dutch_univ}(b). It follows the well-known conflict structure, a rise and fall, in novels: exposition, inciting incident, rising action, climax, falling action, resolution, and denouement~\cite{freytag1908freytag}. At the beginning of the semester, this conflict structure arises, and the maximum value of $Z$-score is around the twelve week at the end of the semester. This motif structure appears less and less remarkably during the vacation (after the twelve week). We guess that the conflict may be resolved indeed or fade out as the students move away from each other. In the inset, the motif ignoring sign does not show this tendency. The negative value of $Z$-score for the motif 16 means that the motif itself are rarely discovered as compared with other motifs because the negative links are naturally rare. However, the result in Fig.~\ref{fig:dutch_univ}(b) tells us that some motifs have their own epic, which can be easy to be ignored but sometimes give important information, e.g., bullying phenomena at school. The bullying happens in confidence mostly, so the motif being able to portray the bullying is less likely to be observed in a student network. 
Analyzing the bullying motif may help to understand the situation.

\section{Conclusions} \label{sec:conclusions}
In our surroundings, the interactions in various systems can include a positive or negative effect on others. The interaction is represented as a link in complex networks, and the effect is in a form of a sign on the link. In this paper, we have tried to understand how the local interaction structure affects the emergent behavior of a network, and narrowed the question to what motifs mainly devote to the growth of the giant component (network itself, herein). Especially, we have considered the sign and direction of a link, simultaneously.

To quantitatively measure the significant abundance of motif, we have calculated the $Z$-score. We have first investigated the motif dynamics in a synthetic ecological system and found that the closed triplets motifs $m\in\mathbb{M}(3, 3)$ with more than two positive links are correlated with the growth of the network. Here, the non-cyclic motif has the highest abundance which may be related to the stability of the system, and we have confirmed the higher abundances of the non-cyclic motifs than that of the cyclic motifs in other empirical data. 

As a further understanding of motif dynamics, we have looked into the dynamics of motif of interest, that is, the conflict motif in the friendship network. The dynamics has been able to be interpreted properly under the circumstances. From this result, we have concluded that some motifs are worth to be deeply analyzed depending on networks, although they do not have significant abundances. 

For future works, we will investigate the relation between the motif abundance and the stability or robustness of the system in more realistic models. If one reveals the direct connection between the abundance of the non-cyclic motifs and the system robustness, it ensures that the real systems evolve towards the higher abundance of the non-cyclic structural motif to enhance the stability of their whole structure. It also leads to the new design and proposal of growing successful strategies of networks.

\section{Acknowledgments} \label{sec:acknowledgements}

This work was supported by the National Research Foundation (NRF) of Korea through the Grant No. NRF-2020R1A2C2010875 funded by the Korea government.

%


\begin{thebibliography}{34}%
\makeatletter
\providecommand \@ifxundefined [1]{%
 \@ifx{#1\undefined}
}%
\providecommand \@ifnum [1]{%
 \ifnum #1\expandafter \@firstoftwo
 \else \expandafter \@secondoftwo
 \fi
}%
\providecommand \@ifx [1]{%
 \ifx #1\expandafter \@firstoftwo
 \else \expandafter \@secondoftwo
 \fi
}%
\providecommand \natexlab [1]{#1}%
\providecommand \enquote  [1]{``#1''}%
\providecommand \bibnamefont  [1]{#1}%
\providecommand \bibfnamefont [1]{#1}%
\providecommand \citenamefont [1]{#1}%
\providecommand \href@noop [0]{\@secondoftwo}%
\providecommand \href [0]{\begingroup \@sanitize@url \@href}%
\providecommand \@href[1]{\@@startlink{#1}\@@href}%
\providecommand \@@href[1]{\endgroup#1\@@endlink}%
\providecommand \@sanitize@url [0]{\catcode `\\12\catcode `\$12\catcode
  `\&12\catcode `\#12\catcode `\^12\catcode `\_12\catcode `\%12\relax}%
\providecommand \@@startlink[1]{}%
\providecommand \@@endlink[0]{}%
\providecommand \url  [0]{\begingroup\@sanitize@url \@url }%
\providecommand \@url [1]{\endgroup\@href {#1}{\urlprefix }}%
\providecommand \urlprefix  [0]{URL }%
\providecommand \Eprint [0]{\href }%
\providecommand \doibase [0]{http://dx.doi.org/}%
\providecommand \selectlanguage [0]{\@gobble}%
\providecommand \bibinfo  [0]{\@secondoftwo}%
\providecommand \bibfield  [0]{\@secondoftwo}%
\providecommand \translation [1]{[#1]}%
\providecommand \BibitemOpen [0]{}%
\providecommand \bibitemStop [0]{}%
\providecommand \bibitemNoStop [0]{.\EOS\space}%
\providecommand \EOS [0]{\spacefactor3000\relax}%
\providecommand \BibitemShut  [1]{\csname bibitem#1\endcsname}%
\let\auto@bib@innerbib\@empty
\bibitem [{\citenamefont {Paranjape}\ \emph {et~al.}(2017)\citenamefont
  {Paranjape}, \citenamefont {Benson},\ and\ \citenamefont
  {Leskovec}}]{paranjape2017motifs}%
  \BibitemOpen
  \bibfield  {author} {\bibinfo {author} {\bibfnamefont {A.}~\bibnamefont
  {Paranjape}}, \bibinfo {author} {\bibfnamefont {A.~R.}\ \bibnamefont
  {Benson}}, \ and\ \bibinfo {author} {\bibfnamefont {J.}~\bibnamefont
  {Leskovec}},\ }in\ \href@noop {} {\emph {\bibinfo {booktitle} {Proceedings of
  the Tenth ACM International Conference on Web Search and Data Mining}}}\
  (\bibinfo {year} {2017})\ pp.\ \bibinfo {pages} {601--610}\BibitemShut
  {NoStop}%
\bibitem [{\citenamefont {Li}\ \emph {et~al.}(2017)\citenamefont {Li},
  \citenamefont {Cornelius}, \citenamefont {Liu}, \citenamefont {Wang},\ and\
  \citenamefont {Barab{\'a}si}}]{li2017fundamental}%
  \BibitemOpen
  \bibfield  {author} {\bibinfo {author} {\bibfnamefont {A.}~\bibnamefont
  {Li}}, \bibinfo {author} {\bibfnamefont {S.~P.}\ \bibnamefont {Cornelius}},
  \bibinfo {author} {\bibfnamefont {Y.-Y.}\ \bibnamefont {Liu}}, \bibinfo
  {author} {\bibfnamefont {L.}~\bibnamefont {Wang}}, \ and\ \bibinfo {author}
  {\bibfnamefont {A.-L.}\ \bibnamefont {Barab{\'a}si}},\ }\href@noop {}
  {\bibfield  {journal} {\bibinfo  {journal} {Science}\ }\textbf {\bibinfo
  {volume} {358}},\ \bibinfo {pages} {1042} (\bibinfo {year}
  {2017})}\BibitemShut {NoStop}%
\bibitem [{\citenamefont {Cencetti}\ \emph {et~al.}(2020)\citenamefont
  {Cencetti}, \citenamefont {Battiston}, \citenamefont {Lepri},\ and\
  \citenamefont {Karsai}}]{cencetti2020temporal}%
  \BibitemOpen
  \bibfield  {author} {\bibinfo {author} {\bibfnamefont {G.}~\bibnamefont
  {Cencetti}}, \bibinfo {author} {\bibfnamefont {F.}~\bibnamefont {Battiston}},
  \bibinfo {author} {\bibfnamefont {B.}~\bibnamefont {Lepri}}, \ and\ \bibinfo
  {author} {\bibfnamefont {M.}~\bibnamefont {Karsai}},\ }\href@noop {}
  {\bibfield  {journal} {\bibinfo  {journal} {arXiv preprint arXiv:2010.03404}\
  } (\bibinfo {year} {2020})}\BibitemShut {NoStop}%
\bibitem [{\citenamefont {Benson}\ \emph {et~al.}(2016)\citenamefont {Benson},
  \citenamefont {Gleich},\ and\ \citenamefont {Leskovec}}]{benson2016higher}%
  \BibitemOpen
  \bibfield  {author} {\bibinfo {author} {\bibfnamefont {A.~R.}\ \bibnamefont
  {Benson}}, \bibinfo {author} {\bibfnamefont {D.~F.}\ \bibnamefont {Gleich}},
  \ and\ \bibinfo {author} {\bibfnamefont {J.}~\bibnamefont {Leskovec}},\
  }\href@noop {} {\bibfield  {journal} {\bibinfo  {journal} {Science}\ }\textbf
  {\bibinfo {volume} {353}},\ \bibinfo {pages} {163} (\bibinfo {year}
  {2016})}\BibitemShut {NoStop}%
\bibitem [{\citenamefont {Peel}\ \emph {et~al.}(2018)\citenamefont {Peel},
  \citenamefont {Delvenne},\ and\ \citenamefont
  {Lambiotte}}]{peel2018multiscale}%
  \BibitemOpen
  \bibfield  {author} {\bibinfo {author} {\bibfnamefont {L.}~\bibnamefont
  {Peel}}, \bibinfo {author} {\bibfnamefont {J.-C.}\ \bibnamefont {Delvenne}},
  \ and\ \bibinfo {author} {\bibfnamefont {R.}~\bibnamefont {Lambiotte}},\
  }\href@noop {} {\bibfield  {journal} {\bibinfo  {journal} {Proceedings of the
  National Academy of Sciences}\ }\textbf {\bibinfo {volume} {115}},\ \bibinfo
  {pages} {4057} (\bibinfo {year} {2018})}\BibitemShut {NoStop}%
\bibitem [{\citenamefont {Kirkley}\ \emph {et~al.}(2019)\citenamefont
  {Kirkley}, \citenamefont {Cantwell},\ and\ \citenamefont
  {Newman}}]{kirkley2019balance}%
  \BibitemOpen
  \bibfield  {author} {\bibinfo {author} {\bibfnamefont {A.}~\bibnamefont
  {Kirkley}}, \bibinfo {author} {\bibfnamefont {G.~T.}\ \bibnamefont
  {Cantwell}}, \ and\ \bibinfo {author} {\bibfnamefont {M.}~\bibnamefont
  {Newman}},\ }\href@noop {} {\bibfield  {journal} {\bibinfo  {journal}
  {Physical Review E}\ }\textbf {\bibinfo {volume} {99}},\ \bibinfo {pages}
  {012320} (\bibinfo {year} {2019})}\BibitemShut {NoStop}%
\bibitem [{\citenamefont {Itzkovitz}\ \emph {et~al.}(2003)\citenamefont
  {Itzkovitz}, \citenamefont {Milo}, \citenamefont {Kashtan}, \citenamefont
  {Ziv},\ and\ \citenamefont {Alon}}]{itzkovitz2003subgraphs}%
  \BibitemOpen
  \bibfield  {author} {\bibinfo {author} {\bibfnamefont {S.}~\bibnamefont
  {Itzkovitz}}, \bibinfo {author} {\bibfnamefont {R.}~\bibnamefont {Milo}},
  \bibinfo {author} {\bibfnamefont {N.}~\bibnamefont {Kashtan}}, \bibinfo
  {author} {\bibfnamefont {G.}~\bibnamefont {Ziv}}, \ and\ \bibinfo {author}
  {\bibfnamefont {U.}~\bibnamefont {Alon}},\ }\href@noop {} {\bibfield
  {journal} {\bibinfo  {journal} {Physical review E}\ }\textbf {\bibinfo
  {volume} {68}},\ \bibinfo {pages} {026127} (\bibinfo {year}
  {2003})}\BibitemShut {NoStop}%
\bibitem [{\citenamefont {Prill}\ \emph {et~al.}(2005)\citenamefont {Prill},
  \citenamefont {Iglesias},\ and\ \citenamefont
  {Levchenko}}]{prill2005dynamic}%
  \BibitemOpen
  \bibfield  {author} {\bibinfo {author} {\bibfnamefont {R.~J.}\ \bibnamefont
  {Prill}}, \bibinfo {author} {\bibfnamefont {P.~A.}\ \bibnamefont {Iglesias}},
  \ and\ \bibinfo {author} {\bibfnamefont {A.}~\bibnamefont {Levchenko}},\
  }\href@noop {} {\bibfield  {journal} {\bibinfo  {journal} {PLoS Biol}\
  }\textbf {\bibinfo {volume} {3}},\ \bibinfo {pages} {e343} (\bibinfo {year}
  {2005})}\BibitemShut {NoStop}%
\bibitem [{\citenamefont {Grilli}\ \emph {et~al.}(2017)\citenamefont {Grilli},
  \citenamefont {Barab{\'a}s}, \citenamefont {Michalska-Smith},\ and\
  \citenamefont {Allesina}}]{grilli2017higher}%
  \BibitemOpen
  \bibfield  {author} {\bibinfo {author} {\bibfnamefont {J.}~\bibnamefont
  {Grilli}}, \bibinfo {author} {\bibfnamefont {G.}~\bibnamefont {Barab{\'a}s}},
  \bibinfo {author} {\bibfnamefont {M.~J.}\ \bibnamefont {Michalska-Smith}}, \
  and\ \bibinfo {author} {\bibfnamefont {S.}~\bibnamefont {Allesina}},\
  }\href@noop {} {\bibfield  {journal} {\bibinfo  {journal} {Nature}\ }\textbf
  {\bibinfo {volume} {548}},\ \bibinfo {pages} {210} (\bibinfo {year}
  {2017})}\BibitemShut {NoStop}%
\bibitem [{\citenamefont {Luce}\ and\ \citenamefont
  {Perry}(1949)}]{luce1949method}%
  \BibitemOpen
  \bibfield  {author} {\bibinfo {author} {\bibfnamefont {R.~D.}\ \bibnamefont
  {Luce}}\ and\ \bibinfo {author} {\bibfnamefont {A.~D.}\ \bibnamefont
  {Perry}},\ }\href@noop {} {\bibfield  {journal} {\bibinfo  {journal}
  {Psychometrika}\ }\textbf {\bibinfo {volume} {14}},\ \bibinfo {pages} {95}
  (\bibinfo {year} {1949})}\BibitemShut {NoStop}%
\bibitem [{\citenamefont {Watts}\ and\ \citenamefont
  {Strogatz}(1998)}]{watts1998collective}%
  \BibitemOpen
  \bibfield  {author} {\bibinfo {author} {\bibfnamefont {D.~J.}\ \bibnamefont
  {Watts}}\ and\ \bibinfo {author} {\bibfnamefont {S.~H.}\ \bibnamefont
  {Strogatz}},\ }\href@noop {} {\bibfield  {journal} {\bibinfo  {journal}
  {nature}\ }\textbf {\bibinfo {volume} {393}},\ \bibinfo {pages} {440}
  (\bibinfo {year} {1998})}\BibitemShut {NoStop}%
\bibitem [{\citenamefont {Onnela}\ \emph {et~al.}(2005)\citenamefont {Onnela},
  \citenamefont {Saram{\"a}ki}, \citenamefont {Kert{\'e}sz},\ and\
  \citenamefont {Kaski}}]{onnela2005intensity}%
  \BibitemOpen
  \bibfield  {author} {\bibinfo {author} {\bibfnamefont {J.-P.}\ \bibnamefont
  {Onnela}}, \bibinfo {author} {\bibfnamefont {J.}~\bibnamefont
  {Saram{\"a}ki}}, \bibinfo {author} {\bibfnamefont {J.}~\bibnamefont
  {Kert{\'e}sz}}, \ and\ \bibinfo {author} {\bibfnamefont {K.}~\bibnamefont
  {Kaski}},\ }\href@noop {} {\bibfield  {journal} {\bibinfo  {journal}
  {Physical Review E}\ }\textbf {\bibinfo {volume} {71}},\ \bibinfo {pages}
  {065103} (\bibinfo {year} {2005})}\BibitemShut {NoStop}%
\bibitem [{\citenamefont {Shimada}(2014)}]{shimada2014universal}%
  \BibitemOpen
  \bibfield  {author} {\bibinfo {author} {\bibfnamefont {T.}~\bibnamefont
  {Shimada}},\ }\href@noop {} {\bibfield  {journal} {\bibinfo  {journal}
  {Scientific reports}\ }\textbf {\bibinfo {volume} {4}},\ \bibinfo {pages}
  {4082} (\bibinfo {year} {2014})}\BibitemShut {NoStop}%
\bibitem [{\citenamefont {Milo}\ \emph {et~al.}(2002)\citenamefont {Milo},
  \citenamefont {Shen-Orr}, \citenamefont {Itzkovitz}, \citenamefont {Kashtan},
  \citenamefont {Chklovskii},\ and\ \citenamefont {Alon}}]{milo2002network}%
  \BibitemOpen
  \bibfield  {author} {\bibinfo {author} {\bibfnamefont {R.}~\bibnamefont
  {Milo}}, \bibinfo {author} {\bibfnamefont {S.}~\bibnamefont {Shen-Orr}},
  \bibinfo {author} {\bibfnamefont {S.}~\bibnamefont {Itzkovitz}}, \bibinfo
  {author} {\bibfnamefont {N.}~\bibnamefont {Kashtan}}, \bibinfo {author}
  {\bibfnamefont {D.}~\bibnamefont {Chklovskii}}, \ and\ \bibinfo {author}
  {\bibfnamefont {U.}~\bibnamefont {Alon}},\ }\href@noop {} {\bibfield
  {journal} {\bibinfo  {journal} {Science}\ }\textbf {\bibinfo {volume}
  {298}},\ \bibinfo {pages} {824} (\bibinfo {year} {2002})}\BibitemShut
  {NoStop}%
\bibitem [{\citenamefont {Sporns}\ and\ \citenamefont
  {K{\"o}tter}(2004)}]{sporns2004motifs}%
  \BibitemOpen
  \bibfield  {author} {\bibinfo {author} {\bibfnamefont {O.}~\bibnamefont
  {Sporns}}\ and\ \bibinfo {author} {\bibfnamefont {R.}~\bibnamefont
  {K{\"o}tter}},\ }\href@noop {} {\bibfield  {journal} {\bibinfo  {journal}
  {PLoS Biol}\ }\textbf {\bibinfo {volume} {2}},\ \bibinfo {pages} {e369}
  (\bibinfo {year} {2004})}\BibitemShut {NoStop}%
\bibitem [{\citenamefont {Alon}(2007)}]{alon2007network}%
  \BibitemOpen
  \bibfield  {author} {\bibinfo {author} {\bibfnamefont {U.}~\bibnamefont
  {Alon}},\ }\href@noop {} {\bibfield  {journal} {\bibinfo  {journal} {Nature
  Reviews Genetics}\ }\textbf {\bibinfo {volume} {8}},\ \bibinfo {pages} {450}
  (\bibinfo {year} {2007})}\BibitemShut {NoStop}%
\bibitem [{\citenamefont {Facchetti}\ \emph {et~al.}(2011)\citenamefont
  {Facchetti}, \citenamefont {Iacono},\ and\ \citenamefont
  {Altafini}}]{facchetti2011computing}%
  \BibitemOpen
  \bibfield  {author} {\bibinfo {author} {\bibfnamefont {G.}~\bibnamefont
  {Facchetti}}, \bibinfo {author} {\bibfnamefont {G.}~\bibnamefont {Iacono}}, \
  and\ \bibinfo {author} {\bibfnamefont {C.}~\bibnamefont {Altafini}},\
  }\href@noop {} {\bibfield  {journal} {\bibinfo  {journal} {Proceedings of the
  National Academy of Sciences}\ }\textbf {\bibinfo {volume} {108}},\ \bibinfo
  {pages} {20953} (\bibinfo {year} {2011})}\BibitemShut {NoStop}%
\bibitem [{\citenamefont {Kim}\ \emph {et~al.}(2014{\natexlab{a}})\citenamefont
  {Kim}, \citenamefont {Roh}, \citenamefont {Jeong},\ and\ \citenamefont
  {Son}}]{kim2014dynamic}%
  \BibitemOpen
  \bibfield  {author} {\bibinfo {author} {\bibfnamefont {Y.~J.}\ \bibnamefont
  {Kim}}, \bibinfo {author} {\bibfnamefont {M.}~\bibnamefont {Roh}}, \bibinfo
  {author} {\bibfnamefont {S.-Y.}\ \bibnamefont {Jeong}}, \ and\ \bibinfo
  {author} {\bibfnamefont {S.-W.}\ \bibnamefont {Son}},\ }\href@noop {}
  {\bibfield  {journal} {\bibinfo  {journal} {Journal of the Korean Physical
  Society}\ }\textbf {\bibinfo {volume} {65}},\ \bibinfo {pages} {1709}
  (\bibinfo {year} {2014}{\natexlab{a}})}\BibitemShut {NoStop}%
\bibitem [{\citenamefont {Kim}\ \emph {et~al.}(2014{\natexlab{b}})\citenamefont
  {Kim}, \citenamefont {Roh},\ and\ \citenamefont {Son}}]{kim2014network}%
  \BibitemOpen
  \bibfield  {author} {\bibinfo {author} {\bibfnamefont {Y.~J.}\ \bibnamefont
  {Kim}}, \bibinfo {author} {\bibfnamefont {M.}~\bibnamefont {Roh}}, \ and\
  \bibinfo {author} {\bibfnamefont {S.-W.}\ \bibnamefont {Son}},\ }\href@noop
  {} {\bibfield  {journal} {\bibinfo  {journal} {Journal of the Korean Physical
  Society}\ }\textbf {\bibinfo {volume} {64}},\ \bibinfo {pages} {341}
  (\bibinfo {year} {2014}{\natexlab{b}})}\BibitemShut {NoStop}%
\bibitem [{\citenamefont {Murase}\ \emph {et~al.}(2010)\citenamefont {Murase},
  \citenamefont {Shimada},\ and\ \citenamefont {Ito}}]{murase2010simple}%
  \BibitemOpen
  \bibfield  {author} {\bibinfo {author} {\bibfnamefont {Y.}~\bibnamefont
  {Murase}}, \bibinfo {author} {\bibfnamefont {T.}~\bibnamefont {Shimada}}, \
  and\ \bibinfo {author} {\bibfnamefont {N.}~\bibnamefont {Ito}},\ }\href@noop
  {} {\bibfield  {journal} {\bibinfo  {journal} {New Journal of Physics}\
  }\textbf {\bibinfo {volume} {12}},\ \bibinfo {pages} {063021} (\bibinfo
  {year} {2010})}\BibitemShut {NoStop}%
\bibitem [{\citenamefont {Murase}\ \emph {et~al.}(2015)\citenamefont {Murase},
  \citenamefont {Shimada}, \citenamefont {Ito},\ and\ \citenamefont
  {Rikvold}}]{murase2015universal}%
  \BibitemOpen
  \bibfield  {author} {\bibinfo {author} {\bibfnamefont {Y.}~\bibnamefont
  {Murase}}, \bibinfo {author} {\bibfnamefont {T.}~\bibnamefont {Shimada}},
  \bibinfo {author} {\bibfnamefont {N.}~\bibnamefont {Ito}}, \ and\ \bibinfo
  {author} {\bibfnamefont {P.~A.}\ \bibnamefont {Rikvold}},\ }in\ \href@noop {}
  {\emph {\bibinfo {booktitle} {Proceedings of the International Conference on
  Social Modeling and Simulation, plus Econophysics Colloquium 2014}}}\
  (\bibinfo {organization} {Springer, Cham},\ \bibinfo {year} {2015})\ pp.\
  \bibinfo {pages} {175--186}\BibitemShut {NoStop}%
\bibitem [{\citenamefont {Ogushi}\ \emph {et~al.}(2017)\citenamefont {Ogushi},
  \citenamefont {Kert{\'e}sz}, \citenamefont {Kaski},\ and\ \citenamefont
  {Shimada}}]{ogushi2017enhanced}%
  \BibitemOpen
  \bibfield  {author} {\bibinfo {author} {\bibfnamefont {F.}~\bibnamefont
  {Ogushi}}, \bibinfo {author} {\bibfnamefont {J.}~\bibnamefont {Kert{\'e}sz}},
  \bibinfo {author} {\bibfnamefont {K.}~\bibnamefont {Kaski}}, \ and\ \bibinfo
  {author} {\bibfnamefont {T.}~\bibnamefont {Shimada}},\ }\href@noop {}
  {\bibfield  {journal} {\bibinfo  {journal} {Scientific Reports}\ }\textbf
  {\bibinfo {volume} {7}},\ \bibinfo {pages} {1} (\bibinfo {year}
  {2017})}\BibitemShut {NoStop}%
\bibitem [{\citenamefont {Murase}\ and\ \citenamefont
  {Rikvold}(2018)}]{murase2018conservation}%
  \BibitemOpen
  \bibfield  {author} {\bibinfo {author} {\bibfnamefont {Y.}~\bibnamefont
  {Murase}}\ and\ \bibinfo {author} {\bibfnamefont {P.~A.}\ \bibnamefont
  {Rikvold}},\ }\href@noop {} {\bibfield  {journal} {\bibinfo  {journal} {New
  Journal of Physics}\ }\textbf {\bibinfo {volume} {20}},\ \bibinfo {pages}
  {083023} (\bibinfo {year} {2018})}\BibitemShut {NoStop}%
\bibitem [{\citenamefont {Ogushi}\ \emph {et~al.}(2019)\citenamefont {Ogushi},
  \citenamefont {Kert{\'e}sz}, \citenamefont {Kaski},\ and\ \citenamefont
  {Shimada}}]{ogushi2019temporal}%
  \BibitemOpen
  \bibfield  {author} {\bibinfo {author} {\bibfnamefont {F.}~\bibnamefont
  {Ogushi}}, \bibinfo {author} {\bibfnamefont {J.}~\bibnamefont {Kert{\'e}sz}},
  \bibinfo {author} {\bibfnamefont {K.}~\bibnamefont {Kaski}}, \ and\ \bibinfo
  {author} {\bibfnamefont {T.}~\bibnamefont {Shimada}},\ }\href@noop {}
  {\bibfield  {journal} {\bibinfo  {journal} {Royal Society open science}\
  }\textbf {\bibinfo {volume} {6}},\ \bibinfo {pages} {181471} (\bibinfo {year}
  {2019})}\BibitemShut {NoStop}%
\bibitem [{\citenamefont {Kunegis}(2013)}]{konect}%
  \BibitemOpen
  \bibfield  {author} {\bibinfo {author} {\bibfnamefont {J.}~\bibnamefont
  {Kunegis}},\ }in\ \href
  {http://userpages.uni-koblenz.de/~kunegis/paper/kunegis-koblenz-network-collection.pdf}
  {\emph {\bibinfo {booktitle} {Proc. Int. Conf. on World Wide Web
  Companion}}}\ (\bibinfo {year} {2013})\ pp.\ \bibinfo {pages}
  {1343--1350}\BibitemShut {NoStop}%
\bibitem [{\citenamefont {Iorio}\ \emph {et~al.}(2016)\citenamefont {Iorio},
  \citenamefont {Bernardo-Faura}, \citenamefont {Gobbi}, \citenamefont
  {Cokelaer}, \citenamefont {Jurman},\ and\ \citenamefont
  {Saez-Rodriguez}}]{iorio2016efficient}%
  \BibitemOpen
  \bibfield  {author} {\bibinfo {author} {\bibfnamefont {F.}~\bibnamefont
  {Iorio}}, \bibinfo {author} {\bibfnamefont {M.}~\bibnamefont
  {Bernardo-Faura}}, \bibinfo {author} {\bibfnamefont {A.}~\bibnamefont
  {Gobbi}}, \bibinfo {author} {\bibfnamefont {T.}~\bibnamefont {Cokelaer}},
  \bibinfo {author} {\bibfnamefont {G.}~\bibnamefont {Jurman}}, \ and\ \bibinfo
  {author} {\bibfnamefont {J.}~\bibnamefont {Saez-Rodriguez}},\ }\href@noop {}
  {\bibfield  {journal} {\bibinfo  {journal} {BMC bioinformatics}\ }\textbf
  {\bibinfo {volume} {17}},\ \bibinfo {pages} {542} (\bibinfo {year}
  {2016})}\BibitemShut {NoStop}%
\bibitem [{\citenamefont {Lidicker~Jr}(1979)}]{lidicker1979clarification}%
  \BibitemOpen
  \bibfield  {author} {\bibinfo {author} {\bibfnamefont {W.~Z.}\ \bibnamefont
  {Lidicker~Jr}},\ }\href@noop {} {\bibfield  {journal} {\bibinfo  {journal}
  {Bioscience}\ }\textbf {\bibinfo {volume} {29}},\ \bibinfo {pages} {475}
  (\bibinfo {year} {1979})}\BibitemShut {NoStop}%
\bibitem [{\citenamefont {Sahney}\ \emph {et~al.}(2010)\citenamefont {Sahney},
  \citenamefont {Benton},\ and\ \citenamefont {Ferry}}]{sahney2010links}%
  \BibitemOpen
  \bibfield  {author} {\bibinfo {author} {\bibfnamefont {S.}~\bibnamefont
  {Sahney}}, \bibinfo {author} {\bibfnamefont {M.~J.}\ \bibnamefont {Benton}},
  \ and\ \bibinfo {author} {\bibfnamefont {P.~A.}\ \bibnamefont {Ferry}},\
  }\href@noop {} {\bibfield  {journal} {\bibinfo  {journal} {Biology letters}\
  }\textbf {\bibinfo {volume} {6}},\ \bibinfo {pages} {544} (\bibinfo {year}
  {2010})}\BibitemShut {NoStop}%
\bibitem [{\citenamefont {Park}\ \emph {et~al.}(2020)\citenamefont {Park},
  \citenamefont {Pichugin},\ and\ \citenamefont {Traulsen}}]{park2020cyclic}%
  \BibitemOpen
  \bibfield  {author} {\bibinfo {author} {\bibfnamefont {H.~J.}\ \bibnamefont
  {Park}}, \bibinfo {author} {\bibfnamefont {Y.}~\bibnamefont {Pichugin}}, \
  and\ \bibinfo {author} {\bibfnamefont {A.}~\bibnamefont {Traulsen}},\
  }\href@noop {} {\bibfield  {journal} {\bibinfo  {journal} {bioRxiv}\ }
  (\bibinfo {year} {2020})}\BibitemShut {NoStop}%
\bibitem [{\citenamefont {Van~de Bunt}\ \emph {et~al.}(1999)\citenamefont
  {Van~de Bunt}, \citenamefont {Van~Duijn},\ and\ \citenamefont
  {Snijders}}]{van1999friendship}%
  \BibitemOpen
  \bibfield  {author} {\bibinfo {author} {\bibfnamefont {G.~G.}\ \bibnamefont
  {Van~de Bunt}}, \bibinfo {author} {\bibfnamefont {M.~A.}\ \bibnamefont
  {Van~Duijn}}, \ and\ \bibinfo {author} {\bibfnamefont {T.~A.}\ \bibnamefont
  {Snijders}},\ }\href@noop {} {\bibfield  {journal} {\bibinfo  {journal}
  {Computational and Mathematical Organization Theory}\ }\textbf {\bibinfo
  {volume} {5}},\ \bibinfo {pages} {167} (\bibinfo {year} {1999})}\BibitemShut
  {NoStop}%
\bibitem [{kon(2016{\natexlab{a}})}]{konect:2016:chess}%
  \BibitemOpen
  \href {http://konect.uni-koblenz.de/networks/chess} {\enquote {\bibinfo
  {title} {Chess network dataset -- {KONECT}},}\ } (\bibinfo {year}
  {2016}{\natexlab{a}})\BibitemShut {NoStop}%
\bibitem [{kon(2016{\natexlab{b}})}]{konect:2016:elec}%
  \BibitemOpen
  \href {http://konect.uni-koblenz.de/networks/elec} {\enquote {\bibinfo
  {title} {Wikipedia elections network dataset -- {KONECT}},}\ } (\bibinfo
  {year} {2016}{\natexlab{b}})\BibitemShut {NoStop}%
\bibitem [{\citenamefont {Leskovec}\ \emph {et~al.}(2010)\citenamefont
  {Leskovec}, \citenamefont {Huttenlocher},\ and\ \citenamefont
  {Kleinberg}}]{konect:leskovec207}%
  \BibitemOpen
  \bibfield  {author} {\bibinfo {author} {\bibfnamefont {J.}~\bibnamefont
  {Leskovec}}, \bibinfo {author} {\bibfnamefont {D.}~\bibnamefont
  {Huttenlocher}}, \ and\ \bibinfo {author} {\bibfnamefont {J.}~\bibnamefont
  {Kleinberg}},\ }in\ \href@noop {} {\emph {\bibinfo {booktitle} {Proc. Int.
  Conf. on Weblogs and Social Media}}}\ (\bibinfo {year} {2010})\BibitemShut
  {NoStop}%
\bibitem [{\citenamefont {Freytag}\ and\ \citenamefont
  {MacEwan}(1908)}]{freytag1908freytag}%
  \BibitemOpen
  \bibfield  {author} {\bibinfo {author} {\bibfnamefont {G.}~\bibnamefont
  {Freytag}}\ and\ \bibinfo {author} {\bibfnamefont {E.~J.}\ \bibnamefont
  {MacEwan}},\ }\href@noop {} {\emph {\bibinfo {title} {Freytag's technique of
  the drama: an exposition of dramatic composition and art}}}\ (\bibinfo
  {publisher} {Scott, Foresman and Company},\ \bibinfo {year}
  {1908})\BibitemShut {NoStop}%
\end{thebibliography}

\end{document}